\documentclass[12pt,preprint]{aastex}

\received{2006 March 10}
\begin{document}
\def\pedant{Mart{\'{\i}}n }
\def\etal{{\sl et al.}}
\def\etall{{\sl et al. }}
\def\pma{$\arcsec$~yr$^{-1}$ }
\def\kms{km~s$^{-1}$ }
\def\msun{$M_{\odot}$}
\def\rsun{$R_{\odot}$}
\def\lsun{$L_{\odot}$}
\def\halpha{H$\alpha$}
\def\hbeta{H$\beta$}
\def\hgama{H$\gamma$}
\def\hdelta{H$\delta$}
\def\Teff{T$_{eff}$}
\def\logg{$log_g$} 

\title{A search for binary systems among the nearest L dwarfs\altaffilmark{1}}

\author{I. Neill Reid, E. Lewitus}
\affil{Space Telescope Science Institute,
3700 San Marin Drive, Baltimore, MD 21218, inr@stsci.edu}

\author{ P. R. Allen}
\affil{Department of Astrophysics, Pennsylvania State University, 525 Davey Lab, University Park, Pennsylvania 16801, pallen@astro.psu.edu}

\author{ Kelle L. Cruz\altaffilmark{2}}
\affil{Department of Astrophysics, American Museum of Natural History, Central Park West at 79th St., New York, NY 10024, kelle@amnh.org }

\author{ Adam J. Burgasser}
\affil{ Massachusetts Institute of Technology, Kavli Institute for Astrophysics 
and Space Research, Building 37, Room 664B, 77 Massachusetts Avenue, Cambridge, MA 02139; ajb@mit.edu}

\altaffiltext{1}{ Based on observations made with the NASA/ESA Hubble Space Telescope, obtained from the Space Telescope Science Institute, which is operated by the Association of Universities for Research in Astronomy, Inc., under NASA contract NAS 5-26555. }
\altaffiltext{2}{NSF Astronomy and Astrophysics Postdoctoral Fellow}

\begin{abstract}     
We have used the NICMOS NIC1 camera on the Hubble Space Telescope to obtain high angular resolution images of 52 ultracool dwarfs in the immediate Solar Neighbourhood. Nine systems are resolved as binary, with component separations from 1.5 and 15 AU. Based on current theoretical models and empirical bolometric corrections, all systems have components with similar luminosities, and, consequently, high mass ratios, $q > 0.8$. Limiting analysis to L dwarfs within 20 parsecs, the observed binary fraction is $12^{+7}_{-3}\%$. Applying Bayesian analysis to our dataset, we derive a mass-ratio distribution that peaks strongly at unity. Modelling the semi-major axis distribution as a logarithmic Gaussian, the best fit is centered at log$a_0 = 0.8$ AU ($\sim6.3$ AU), with a (logarithmic) width of $\pm0.3$. The current data are consistent with an overall binary frequency of $\sim24\%$. 
\end{abstract}


\section{Introduction}

Over the last four years we have been undertaking a census of the lower-mass constituents of the immediate Solar Neighborhood (Reid \& Cruz, 2002 - Paper I), concentrating, in particular, on ultracool dwarfs (spectral types M7 and later) within 20 parsecs of the Sun (Cruz et al, 2003; Cruz et al, 2006; Reid et al, in prep.). As part of that survey, we have compiled an all-sky catalogue of 87 L dwarfs in 80 systems with formal distance estimates less than 20 parsecs. This sample offers an opportunity to investigate the statistical characteristics of the local L dwarf population.

Binarity is a key property of low-mass stars and brown dwarfs. Both the overall frequency of binary systems and the distribution of their properties (particularly mass ratios, $q$, and separations, $\Delta$) have emerged as potential tests of various formation theories. Ultracool dwarfs have been the targets of numerous high-resolution imaging surveys, both using adaptive optics on ground-based telescopes and with the Hubble Space Telescope (HST). As summarised most recently by Burgasser et al (2006), the results of those surveys indicate an observed frequency of $\sim15\%$; this compares with an overall binary frequency of 30 to 40\% for M dwarfs (Fischer \& Marcy, 1992; Reid \& Gizis, 1997) and 60 to 70\% for G dwarfs (Duquennoy \& Mayor, 1991). The overwhelming majority of ultracool binaries have small separations, $\Delta < 15$ AU. The nearest ultracool dwarfs are therefore the prime targets for multiplicity surveys, since they provide the optimal resolution in linear units. Those systems also provide the best sensitivities for detecting very low luminosity companions, although there is growing evidence for a tendency towards mass ratios close to unity among ultracool binaries. 

We have been using the NICMOS camera on HST to search for binary systems among the nearest L dwarfs. To date, we have acquired observations of 52 ultracool systems, of which 49 are classed as spectral type L and three as late-type M dwarfs; nine are resolved as binaries, including the L/T system 2MASSW J22521073-1730134 (Reid et al, 2006 - hereinafter RLBC06). High spatial-resolution images are available from the literature for a further 4 L-dwarf systems within 20 parsecs. We present our observations in the following section; the characteristics of the candidate binaries are discussed in \S3; we consider the statistical properties of the full dataset, and the implications in \S4; the final section summarises our results and conclusions.

\section {Observations}

The nearby late-M and L dwarfs imaged in this program were observed as part of a Cycle 13 HST SNAPSHOT program. Most of these systems have spectroscopic distance estimates, and while all were placed within 20 parsecs at the outset of our observational program, a handful have revised distances that lie beyond that limit. All targets were observed with the NIC1 camera and the F110W and F170M filters using the same exposure sequences. The observations in both filters consist of a pair MULTIACCUM exposures, nodding 2.0 arcseconds between the two exposures. The total exposure times are 284 seconds at F110W and 896 seconds in the F170M filter. As discussed in \S2.2, the combined data give limiting magnitudes of m$_{110} \sim 21.9$ and m$_{170} \sim 20.0$ magnitudes (on a Vegamag system) for isolated point sources. Late-type T dwarfs have (m$_{110}$ - m$_{170}$) colours of $\sim1.3$ magnitudes; thus, the F110W data offer the highest sensitivity for the detection of faint companions to the targeted L dwarfs.

\subsection {Identifying binary systems}

The NICMOS data were processed through the standard HST pipeline, and we have analysed the final mosaiced image using standard {\sl IRAF} routines\footnote{IRAF is distributed by the National Optical Astronomy Observatories, which are operated by the Association of Universities for Research in Astronomy, Inc., under cooperative agreement with the National Science Foundation.}. The NIC1 data have a plate-scale of 0.043 arcseconds pixel$^{-1}$, while the formal resolution of the HST data is 0.09 arcseconds with the F110W filter and 0.14 arcseconds with the F170M filter. Since the F110W images have both higher angular resolution and better sensitivity, we have concentrated on those data in our search for faint companions.

We have searched for potential binary companions using a variety of techniques. First, visual inspection of the images reveals a number of systems with obvious close companions. Second, we have used the {\sl imexam} and {\sl daophot} routines in {\sl IRAF} to measure the point spread function (psf), searching for sources with broad full-width half-maxima or unusual profiles. Finally, as is evident from the images of the candidate binaries, unresolved point sources possess a strong Airy disk, with the flux level rising to $\sim10\%$ of the peak flux at radial separations of 0.205 arcseconds in the F110W data and 0.24 arcseconds at F170M. This obviously affects the potential detection of companions at those radii. 

We have analysed these data using the same techniques outlined in our discussion of the L/T binary, 2M2252-1730 (RLBC06). One of the sources observed in our program is 2M0825+2115. This L7.5 dwarf has previous HST observations with WFPC2 (Reid et al, 2001), which demonstrate that the image profile matches a single point source at optical wavelengths. We have therefore taken this object as the psf template for the NIC1 observations\footnote{In RLBC06 we used our data for 2M0045+1634 as the template; this source is unresolved by NICMOS, and subtracting the 2M0825+2115 data shows no evidence for any significant residuals.}. We have used the {\sl imshift} IRAF routine to align the 2M0825 image with each of the other L dwarf targets; scaled the reference data to match the peak flux; and subtracted the 2M0825 data, leaving a `cleaned' image of the environs of the target. There are imperfections in most subtractions, since the NICMOS psf profile changes on relatively short timescales, but none of the low-level residuals have profiles resembling a very faint companion. 

Based on our analysis, forty-three ultracool dwarfs show no significant evidence for binarity. Most have psf profiles with full-width half-maxima (FWHM) of 2.3 to 2.4 pixels (0.099 to 0.103 arcseconds). Three L dwarfs have slightly broader profiles: 2M1507-1627 and 2M1936-5502, with FWHM = 2.47 pixels (0.106 arcseconds); and 2M0036+1820, with FWHM = 2.56 (0.110 arcseconds). Subtracting the 2M0825 template psf show no evidence for the presence a secondary component, and the broader profiles are probably an instrumental effect. Pertinent data for the unresolved ultracool dwarfs are given in Table 1. In most cases, the distance estimates rest on the Cruz et al (2003) spectral-type/M$_J$ relation, and therefore have uncertainties of $\sim15\%$.
 
The remaining nine dwarfs observed in this program are identified as probable binaries. Figure 1 presents NICMOS F110W images of seven sources: higher resolution ACS images of the LHS 102BC (GJ 1001BC) system are discussed by Golimowski et al (2004b), while images of the 2M2252 system are presented by RLBC06. Data for all the candidate binaries are given in Table 2. We discuss these systems in more detail in \S3. 

\subsection {Photometry}

We have utilised two techniques to determine instrumental magnitudes from the NICMOS images. First, we used the {\sl phot} routine in {\sl daophot} to determine aperture photometry for well-isolated sources (the `single' objects and candidate binaries with separations exceeding 0.5 arcseconds). In these cases, we adopted an aperture size of radius 9 pixels (0.36 arcseconds). For the close binaries, we use a smaller aperture size, correcting to 9-pixel photometry using aperture corrections derived from measurements of 2M0825+2115. In the latter cases, we have also estimated the relative magnitudes of the two sources by measuring the peak flux of each component using the IRAF {\sl imexam} profile-fitting routine; combining these data with aperture photometry of both components gives the magnitudes of the two components.

Our aperture photometry is tied to the Vega magnitude scale using the standard HST flux calibration and flux zeropoints of 1786 Janskys and 946 Janskys at F110W and F170M, respectively. We have also used the results given by Schultz et al (2005) to apply appropriate corrections to adjust our photometry to infinite aperture. The resultant magnitudes, m$_{110}$ and m$_{170}$, are listed in Tables 1 and 2. All of these dwarfs have JHK$_S$ photometry from the 2MASS database; indeed, we have used this photometry to derive color terms between the F110W/F1170M and J/H magnitude systems (see RLBC06). The 2MASS data are also given in Tables 1 and 2, and we have used the relative magnitudes in the HST systems to estimate J and H magnitudes for the individual components of the candidate binary systems. 

\section {Ultracool binaries}

\subsection {The present sample}

We have identified nine ultracool dwarfs as probable binaries. Table 2 gives the observed properties for these systems, and Table 3 lists the intrinsic properties inferred for the individual components. We have computed luminosities for each component by applying J-band bolometric corrections as a function of spectral type. Figure 2 shows the basis for our calibration, plotting data for M, L and T dwarfs from the analysis of Golimowski et al (2004a). As discussed further in the following section, we estimate masses from the M$_{bol}$ estimated using the models computed by Burrows et al (1997).

The probable (or confirmed) binary systems are as follows:
\begin{description}

\item[LHS 102BC/GJ 1001BC:] The companion to the nearby M3.5 dwarf LHS 102 was discovered originally by Goldman et al (1999), and the L dwarf was itself revealed as double in NICMOS observations obtained as part of a snapshot survey of stars within 10 parsecs of the Sun (Golimowski et al, 2004b). The system is barely resolved with NICMOS, but the binary status was confirmed through optical observations with the Advanced Camera for Surveys. This system is a classical ultracool binary, with near equal-magnitude components. Golimowski et al point out that the L dwarf properties appear inconsistent with the trigonometric parallax of $104\pm11$ milliarcseconds cited by van Altena et al (1995) for the primary, and they suggest that a distance closer to 15 parsecs is more plausible. New trigonometric parallax measurements are currently being undertaken by the CTIOPI consortium, and we refer the interested reader to Golimowski et al (2004b) for further discussion. It is clear that the system lies within 20 parsecs and, for present purposes, we adopt the larger distance in computing the intrinsic parameters listed in Table 3.

\item[2M0025+4759:] Originally classed as L5 based on near-infrared data, optical spectra indicate a type of L4, consistent with a distance of 23 parsecs for a single dwarf. The HST observations resolve the system into two near-equal luminosity components, implying a distance of 31$\pm$7 parsecs. The system has also been resolved in ground-based observations with the Keck Laser Guide Star AO system (Liu, 2006, priv. comm.). 2M0025+4759 lies only $\sim3.5$ arcminutes from HD 2057 (G171-58/G217-47), a solar metallicity F8 dwarf (Carney et al, 1994) with an Hipparcos parallax placing it at a distance of 42$\pm$2 parsecs. HD 2057 itself is likely a close binary (Latham et al, 2002; Balega et al, 2004). Based on matching Str{\"o}mgren photometry against isochrones, Nordstr{\"o}m et al (2004) estimate an age of $\sim1.1$ Gyrs for HD 2057, with an upper limit of 3.6 Gyrs and no specified lower limit. \\ S. Schmidt (2006, priv. comm.) has combined 2MASS near-infrared and POSS II I-band images (5.9 year baseline) to derive proper motions of ($\mu_\alpha, \mu_\delta) = (+0.312\pm0.034, -0.009\pm0.044)$ arcsec yr$^{-1}$ for the L dwarf. Those data are in reasonable agreement with the Hipparcos astrometry of HD 2057, ($\mu_\alpha, \mu_\delta) = (+0.274\pm0.001, +0.011\pm0.001$) arcsec yr$^{-1}$. If 2M0025+4759 {\sl is} a wide companion of HD 2057, it lies at a separation of $\sim8,800$ AU. This would make it the widest known binary with an ultracool component, but the separation lies within the span of other binaries of comparable total mass (see Figure 8 in Reid \& Walkowicz, 2006). Finally, strong lithium absorption is evident in the combined spectrum, indicating that both components are brown dwarfs with $M < 0.065 M_\odot$ (see Schmidt \& Cruz, 2006, for further discussion). Matched against either the Burrows et al (1997) or Baraffe et al (1998) models, the absence of significant lithium depletion implies an age less than 1 Gyr, broadly consistent with the age estimated for HD 2057.

\item[2M0147-4954:] This dwarf was targeted for observation based on a preliminary spectral type of L0. We have since revised the classification to M8, pushing the system beyond the 20-parsec limit even as a single dwarf; as a binary, with likely spectral types of M8 and L2, we estimate the distance as $\sim33$ parsecs. The flux ratio of the components is similar to that in the 2M0429-3123 system, and the Burrows et al (1997) models indicate a similar mass ratio.

\item[2M0429-3123:] This M7.5 dwarf was resolved originally by Siegler et al (2005) using adaptive optics on the ESO VLT. We estimate a distance of 11.5 parsecs based on the J magnitude of the primary (J$_1$) and the spectral-type/M$_J$ calibration from Cruz et al (2003). There are no indications that the system is particularly young. Siegler et al estimate a spectral type of L1 for the secondary.

\item[2M0700+3157:] This is the only L dwarf in the present sample with a direct trigonometric parallax measurement (Thorstensen \& Kirkpatrick, 2003). With M$_J \sim 14$, the secondary is probably spectral type $\sim$L6.

\item[2M0915+0422:] This ultracool dwarf has been identified independently as a binary system through ground-based AO observations (M. Liu, 2006, priv. comm.). Like LHS 102BC, the two components are almost equal in magnitude and therefore have identical masses. Both components are likely to be brown dwarfs.

\item[2M1707-0558:] This system was first resolved {\sl via} ground-based observations with Spex on the IRTF by Burgasser et al (2004).  Originally classed as spectral type L1 based on the combined optical spectrum, McElwain \& Burgasser (2006) have obtained resolved NIR spectroscopy of this system and derive spectral typee of M9/L0 and $\sim$L3. Their observations also confirm that the components have common proper motion.

\item[2M2152+0937:] This is another equal-luminosity/equal-mass ultracool binary. As with 2M0147-4954, the identification of this dwarf as a binary system removes it from the 20-parsec sample. 

\item[2M2252-1730:] This is one of the handful of L/T binary systems currently known. As discussed in RLBC06, the secondary is noticeably fainter with respect to the primary in the F170M filter than in the F110W passband. Infrared spectroscopy confirms that this is due to the presence of significant methane absorption. Both components are likely to be of substellar mass. 

\end{description}

In most of the observations, the targeted L dwarf is the only object visible in the NICMOS image. There are seven cases, however, where other point sources are visible in the $\sim10 \times 10$ arcsecond NIC1 field of view. With one exception, these sources lie more than 2 arcseconds from the L dwarf and are either bright (J$<16$) and detected in ground-based observations, or extremely faint ($J>19$). These candidate wide companions all have colours and magnitudes that are inconsistent with very low-mass ultracool dwarfs. 

The exception is 2M1705-0516. As Figure 1 shows, this L0.5 dwarf has a faint candidate companion at a separation of 1.36 arcseconds and PA=-5$^o$. The object is unresolved (FWHM$\sim0.10$ arcseconds at F110W), and has m$_{110}=18.00$ and m$_{170} = 16.76$, corresponding to J$\sim17.4$, (J-H)$\sim0.7$. With a galactic latitude of $b \sim +20^o$, this is unlikely to be a reddened background source. The (J-H) colours are consistent with either a mid-type M dwarf at a distance of 1-2 kpc or an early-type T dwarf which, at a distance of 19.5 parsecs, would have M$_J \sim 16.0$. At present, we cannot distinguish between these two possibilities. Follow-up imaging at a later epoch will confirm whether the candidate companion shares the proper motion of the putative primary. For current purposes, we treat 2M1705 as a single ultracool dwarf.

\subsection {Observations of additional systems}

Four L dwarfs from the 20-parsec sample have been observed at high spatial resolution in the course of other binary search programs. These dwarfs are identified in Table 4, where we list relevant data. 

\begin{description}
\item[Denis-P J0205:] One of the three L dwarfs discovered by the DENIS brown dwarf mini-survey (Delfosse et al, 1997), Denis-P J0205 was identified as a binary by Koerner et al (1999) based on K-band imaging with the Keck telescope. Bouy et al (2005) have recently suggested that the brighter component is itself double, and the system consists of two late L dwarfs and a T dwarf.

\item[SDSS 0423-0414:] Originally identified from SDSS observations, this dwarf was classed as type T0 based on its near-infrared spectrum (Geballe et al, 2002). Hawley et al (2002) and Cruz et al (2003) obtained optical spectra in the course of their surveys, and both class the dwarf as type L7.5. Burgasser et al (2005) have resolved the discrepancy; NICMOS observations show that SDSS 0423 is an L/T binary, with properties similar to 2M2252-1730. 

\item[2M0746+2000:] Lying at a distance of $\sim12$ parsecs, this is the brightest L dwarf currently known. Reid et al (2000) originally noted that the dwarf appeared to be overluminous, and high-resolution optical imaging with Wide-Field Camera 2 on HST confirmed that the system is a near equal-magnitude binary (Reid et al, 2001). 

\item[Kelu 1:] The first isolated L dwarf to be discovered (Ruiz, Leggett \& Allard, 1997), trigonometric parallax measurements indicated that Kelu 1 was overluminous, but high-resolution follow-up observations with HST showed no evidence for binarity (Mart\'{\i}n et al, 1999). Those initial observations, however, suffered from bad timing: Liu \& Leggett (2005) have recently resolved the system using ground-based adaptive optics observations, as orbital motion has separated the components. Those observations also resolve a long-standing conundrum: Kelu 1 exhibits weak lithium absorption, implying that, as a single star, it had been caught just as lithium was being depleted; as a binary, these observations are explained due to dilution of the (full-strength) lithium line in component B by continuum from the higher-mass component A. The inferred age for the system is $<800$ Myrs.
\end{description}

All four of these L dwarfs are multiple systems, but this high proportion is not entirely surprising: three are among the brightest L dwarfs known (in apparent magnitude), while the fourth, SDSS 0423, has an unusual spectrum; those properties are correlated directly with binarity, so it is not surprising that these systems were targeted through HST observations. We have therefore not included these systems in our analysis of binary frequency among L dwarfs. A further 31 L dwarfs with formal distances less than 20 parsecs currently lack high resolution imaging data.

\section {Discussion}

\subsection {Companion detection limits}

Our main goal is the detection of low luminosity companions to these ultracool dwarfs. At small separations, the detection limit is set by the psf of the primary; at larger separations, the limit primarily reflects the signal-to-noise of the observations. Figure 3 presents azimuthally-averaged radial profiles in both the F110W and F170M filters for a representative unresolved ultracool dwarf, 2M0523-1403 (m$_{110}=13.66$, m$_{170}=12.32$; J=13.12, H=12.22); the upper panels show the linear profiles (in counts sec$^{-1}$ pix$^{-1}$), and the lower panels plot more extended profiles in a logarithmic scale. 

Figure 3 illustrates both the broader psf in the F170M passband and the higher sensitivity of the F110W imaging. The faintest isolated sources detected in our data have peak count-rates of $\sim0.12$ counts sec$^{-1}$ pix$^{-1}$ in F110W and $\sim0.15$ counts sec$^{-1}$ pix$^{-1}$ in F170M, corresponding to the dashed lines plotted in the lower panels of Figure 3. These peak count-rates correspond to magnitudes of m$_{110}=21.9$ (J$\sim$21.5) and m$_{170}=20.0$ (H$\sim$20.0) for point sources. Extremely cool brown dwarfs are expected to have neutral colours in (J-H), and are therefore easier to detect in the F110W passband.

The effective sky background increases, and the ability to detect a companion decreases correspondingly, within $\sim$1.25 arcsecond of the central star. As noted above, the typical full-width half-maximum of the F110W psf is $\sim$0.10 arcseconds. Only equal-magnitude binaries, such as LHS 102BC, are detectable at such small separation. LHS 102BC is clearly elongated in our images, but effectively marks the small-separation limit of our survey. 

\subsection {The L dwarf binary frequency}

Previous analyses of binarity in ultracool dwarfs (Reid et al, 2001; Gizis et al, 2003; Bouy et al, 2003; Burgasser et al, 2003; Siegler et al, 2005) are based on magnitude-limited samples. We can cast the present analysis in terms of a volume-limited sample, even though we have only observed a subset of that sample. The formal distance limit of the parent ultracool dwarf survey is 20 parsecs; five L dwarf binaries and 41 unresolved objects (including 2M1705-0516) have formal distance estimates within this limit, a binary fraction of $10.9_{-3}^{+6}$\%, where the uncertainties are derived using the formalism outlined by Burgasser et al (2003).

In most cases, however, the distances listed in Tables 1 and 2 are based on spectroscopic parallaxes; those estimates should be corrected for Malmquist bias in statistical analyses. The M$_J$/spectral-type relation from Paper V has a dispersion of $\sim0.35$ magnitudes; this corresponds to an absolute magnitude correction of $\Delta$M$_J = -0.12$ magnitudes, effectively reducing the limit in apparent distance to 19 parsecs. Five binaries and 38 single stars fall within this limit, giving an observed binary fraction of $11.6_{-3}^{+7}$\% for the present sample. This result is formally lower than previous estimates (see Burgasser et al, 2006), although consistent within the (substantial) uncertainties.

\subsection {Masses and mass ratios for the L dwarf binaries}

Our observations measure luminosity ratios for ultracool binaries. Evolutionary effects complicate calculation of the corresponding mass ratios, since brown dwarfs cool and fade at rates that increase with decreasing mass. This is illustrated in Figure 4, where we plot (M$_{bol}$, mass) isochrones for low-mass star/brown dwarf models by Burrows et al (1997) models and by Chabrier et al (2000). Even though the latter "dusty" models (which do not extend below $\sim900$K or $\sim$T6) are a poor match to the colors of late-type L and T dwarfs, the predicted bolometric magnitudes are in reasonable agreement with the Burrows et al dataset. The labeled horizontal lines mark the locations of the binary components listed in Tables 2 and 3, components with spectral types ranging from M7.5 for 2M0429A to $\sim$T2 for 2M2252B.

Figure 4 clearly shows the strong age dependence of brown dwarf masses. Two further points can be made regarding mass ratios of L dwarf binaries. First, old systems must have high mass ratios; thus, a 5-Gyr-old system with an M8 primary (comparable to 2M0429A) and a T2 secondary (like 2M2252B) has components of mass of 0.07 and 0.085$M_\odot$, and a mass ratio of $q \sim 0.82$. Second, at younger ages ($\tau \le 1$ Gyrs), the isochrones have similar slopes through the L-dwarf r\'egime in the (M$_{bol}$, mass) plane; this implies that $q$ decreases with decreasing age. Thus, the same M8/T2 system has $q \sim 0.5$ at $\tau = 1$ Gyr (0.04 \& 0.08 $M_\odot$), and $q \sim 0.38$ at $\tau = 0.3$ Gyrs (0.06 and 0.023 $M_\odot$). 

All binaries in the current sample are field dwarfs. Consequently, the only direct means of constraining mass/age is the presence of lithium absorption.  2M0700+3157 is the only L dwarf with detected lithium absorption, indicating that both components have a mass below 0.065$M_\odot$ and an age less than $\sim1$ Gyr. It is likely that the absence of other lithium detections reflects the relatively low signal-to-noise of the optical spectra; nonetheless, the net result is that we have no direct age estimates for almost all the sample.

Under these circumstances, we must use models to estimate the likely age distribution. Allen et al (2005) have undertaken this type of calculation, basing their analysis on the Burrows et al (1997) evolutionary models. The results depend on the star formation history adopted for the Galactic disk and, to a lesser extent, the form adopted for the underlying mass function. Figure 5 shows the predicted cumulative age distributions for L0, L5 and L6-8 dwarfs for a constant star formation rate and a power-law mass function, $\Psi(M) \propto M^{-1}$. The three distributions are very similar at young ages, with $\sim30\%$ of each sample younger than $\sim1$ Gyrs. The curves diverge at larger ages, with an increasing fraction of older dwarfs at earlier spectral types. Thus, half of the local L0 dwarfs are expected to be younger than $\sim3$ Gyrs, while the 50$^{th}$ percentile mark is reached at age $\sim1.7$ Gyrs for L6=L8 dwarfs.

As a qualitative guide to the mass ratios of the binaries consider here, we have used the Burrows et al (1997) mass-luminosity relations to estimate component masses for ages of $\tau = 1$ and 3 Gyrs. The exceptions are 2M0700 and Kelu 1, where the presence of strong lithium absorption indicates ages less than 1 Gyr. Those data, and the corresponding mass ratios, are listed in Tables 3 and 4. Under these assumptions, all systems have high mass ratios, $q > 0.6$ for $\tau = 1$ Gyr and $q > 0.85$ for $\tau = 3$ Gyrs.

\subsection {The distribution of mass ratios and separations}

All of the L dwarf binaries listed in Tables 2 and 4 have components with relatively high flux ratios. 2M1707AB exhibits the largest magnitude difference, with $\Delta$J = 1.75 magnitudes, and most systems have $\Delta$J$<0.4$ magnitudes. How do these flux ratios compare with the detection limits of the NICMOS observations? 

Figure 6 plots the F110W psf in magnitudes, scaling the measurements relative to the peak brightness. We mark the location of the ultracool companions listed in Tables 2 and 4. The dotted lines mark the effective detection limits spanned by the present set of NICMOS observations. It is clear that all the detected companions are well above those limits. Moreover, as found in previous binary surveys, all of the detected companions lie at relatively small separations. 

To set a rough mass scale for these comparisons, we have used the Burrows et al (1997) models to predict flux ratios for a 0.07$M_\odot$ primary and companions with $0.2 < q < 0.9$ and $\tau = 0.5, 1$ and 5 Gyrs. We choose this value for the primary mass since Allen et al (2005) estimate average masses of $\langle M \rangle = 0.074M_\odot$ for field L0 dwarfs, $0.067M_\odot$ at L5 and $0.063M_\odot$ for L6-L8. High mass (long-lived) brown dwarfs (0.06 to 0.075 $M_\odot$) contribute disproportionately to the local L dwarf population. The resulting flux ratios, shown in Figure 6, suggest that we ought to be able to detect systems with mass ratios as low as $q \sim 0.2$ with the present set of NICMOS observations. As a guide, a $q=0.2$ system comprises an M8 primary (T$_{eff}\sim2440$K, M$_{bol}=13.45$) and $\sim$T7 secondary (T$_{eff}=740$K, M$_{bol}=18.4$) at age 0.5 Gyrs, and  a $\sim$T1 primary (1230K, 16.8) and room-temperature Y-type secondary (350K, 21.8) at age 5 Gyrs\footnote{We assume BC$_J$=2.0 for dwarfs later than T8.}. 

We can quantify our estimates of the underlying mass-ratio and separation distributions through the Bayesian analysis techniques described by Allen et al (2005). Given a particular model for the companion distribution, we can use a disparate set of observations to derive the {\sl posterior distribution}, $P(\theta | D)$, the probability of the model given the data. By Bayes rule, the posterior distribution is the convolution of the likelihood distribution (the likelihood of the data given the model) and the prior distribution (the initial probability of the model). 

Allen et al's (2005) analysis centres on the substellar mass function, but the same techniques can be used to probe the mass function of binary companions. Initial results are included in Burgasser et al's (2006) review, analysing data from previous binary surveys (Koerner et al, 1999; Reid et al, 2001; Gizis et al, 2003; Close et al, 2003; Bouy et al, 2003 and Siegler et al, 2005). In those calculations, the semi-major axis distribution is characterised as a Gaussian in $\log{a}$, with central value $a_0$ and half-width $\sigma_a$, while the mass-ratio distribution is defined as a power-law, index $\gamma$, for a binary fraction $N$. We follow the same approach here, adopting the posterior distribution from the analysis cited in Burgasser et al. (2006) as the prior distribution for our analysis. 

Figure 7 shows the probability distributions derived for each parameter. Expressing $a$ in astronomical units, the best-fit values are $\log{a_0} = 0.8^{+0.06}_{-0.12}$, $\sigma_a = 0.28\pm0.4$, $\gamma = 3.6\pm1$ and $N = 24^{+6}_{-2}\%$. This analysis reinforces the results outlined in Burgasser et al. The best-fit power-law index, $\gamma$, indicates a steep mass function for L dwarf companions, implying a mass-ratio distribution with a strong preference for equal-mass systems. The semi-major axis distribution peaks at $\sim6$ AU, with the model predicting very few systems at separations either beyond $\sim20$ AU or within $\sim1$ AU.

Figure 8 compares the mass-ratio distribution and semi-major axis distribution derived from the present analysis (both the data and the best-fit model) against the results derived by Fischer \& Marcy (1992) for M dwarfs and Duquennoy \& Mayor (1991) for G dwarfs. Clearly, in both cases, the G-dwarf distributions are radically different, while the M dwarf results are closer to our L dwarf analysis. At small separations, imaging data, even with HST, set weaker constraints, leading to the extended tail in the best-fit probability distribution of $a_0$. Maxted \& Jefferies (2005) have argued that significant numbers of spectroscopic binaries remain hidden in L dwarf samples, although their hypothesis currently lacks substantial observational support. With that caveat, our analysis indicates an overall binary fraction of $\sim24\%$, continuing the trend of decreasng binary frequency with decreasing mass.

\section{Summary}

We have presented analysis of high spatial-resolution NICMOS images of 52 ultracool dwarfs in the immediate Solar Neighbourhood. Nine systems are resolved as binary, with component separations from 0.1 to 1.0 arcseconds corresponding to linear separations between 1.5 and 15 AU. Based on current theoretical models and empirical bolometric corrections, all systems have high mass ratios; none includes components with magnitude differences greater than 1.5 magnitudes at J. This is consistent with previous surveys for ultracool binaries in the general field. The observed binary frequency, limiting analysis to stars with Malmquist-corrected distances within 20 parsecs, is of $12_{-3}^{+7}$\%.

Following Allen et al (2005) and Burgasser et al (2006), we have used Bayesian analysis to quantify these results. We derive a mass-ratio distribution that peaks strongly at unity, and matching the semi-major axis distribution with a logarithmic Gaussian gives a best-fit value of $\log{a_0} = 0.8$, or $\sim6.3$ AU. Our analysis indicates that the current data are consistent with an overall L-dwarf binary frequency of $\sim24\%$. 

Acknowledgements: The observations described in this paper are associated with HST program \#10143, and those data were obtained {\sl via} the Hubble Space Telescope data archive facilities maintained at the Space Telescope Science Institute. Support for this research was provided by NASA through a grant from the Space Telescope Science Institute, which is operated by the Association of Universities for Research in Astronomy, Inc., under NASA contract NAS 5-26555. KLC is supported by an NSF Astronomy and Astrophysics Postdoctoral Fellowship under award AST-0401418. \\
This publication makes use of data from the Two Micron All Sky Survey, which is a joint project of the University of Massachusetts and the Infrared Processing and Analysis Center, and funded by the National Aeronautics and Space Administration and the National Science Foundation. 2MASS data were obtained from the NASA/IPAC Infrared Science Archive, which is operated by the Jet Propulsion Laboratory, California Institute of Technology, under contract with the National Aeronautics and Space Administration.


\begin{deluxetable}{rcrrrrrrrc}
\tabletypesize{\scriptsize}
\tablecolumns{10}
\tablenum{1}
\tablewidth{0pt}
\tablecaption{Unresolved ultracool dwarfs}
\tablehead{\colhead{2MASS name} & \colhead{Sp. type} &
\colhead {J} &  \colhead {H} &\colhead {K$_S$} & \colhead{m$_{110}$} & \colhead{m$_{170}$}& \colhead{d (pc.)} & \colhead {Notes }}
\startdata
  2MASS J00361617+1821104 & L3.5 &  12.47 &  11.59 &  11.06 &  13.03 &  11.74 & 8.77$\pm0.06$ & D02, 1 \\
  2MASS J00452143+1634446 & L0  &  13.06 &  12.06 &  11.37 &  13.60 &  12.08 &  18.4$\pm2.8$& 2\\
  2MASS J01075242+0041563 & L8 &  15.82 &  14.51 &  13.71 &  16.53 &  14.65 & 15.6$\pm1.1$ & V04, 3 \\
  2MASS J01235905-4240073 & M8 &  13.15 &  12.47 &  12.04 &  13.60 &  12.58 &  25.1$\pm3.8$& 4   \\
  2MASS J01550354+0950003 & L5 &  14.82 &  13.76 &  13.14 &  15.47 &  13.92 &  18.0$\pm2.7$ & 4  \\
  2MASS J02132880+4444453 &L1.5 & 13.51 & 12.77 & 12.24 & 14.12 & 12.73 & 18.7$\pm2.8$ & 5 \\ 
  2MASS J03140344+1603056 & L0 &  12.53 &  11.82 &  11.24 &  12.97 &  11.85 &  14.4$\pm2.2$ & 4  \\
  2MASS J03552337+1133437 & L6 &  14.05 &  12.53 &  11.53 &  14.74 &  12.64 &  10.1$\pm1.5$& 4   \\
  2MASS J04390101-2353083 & L6.5 &  14.41 &  13.37 &  12.81 &  15.05 &  13.55 &  10.8$\pm1.6$& 5   \\
  2MASS J04455387-3048204 & L2 &  13.41 &  12.57 &  11.98 &  13.96 &  12.64 &  16.6$\pm2.5$& 5   \\
  2MASS J05002100+0330501 & L4 &  13.67 &  12.68 &  12.06 &  14.33 &  12.84 &  13.0$\pm2.0$& 4   \\
  2MASS J05233822-1403022 & L2.5 &  13.12 &  12.22 &  11.63 &  13.66 &  12.32 &  13.4$\pm2.0$& 4   \\
  2MASS J06244595-4521548 & L5 &  14.48 &  13.34 &  12.60 &  15.12 &  13.45 &  15.3$\pm2.3$ & 5   \\
  2MASS J06523073+4710348 & L4.5 &  13.55 &  12.37 &  11.69 &  14.10 &  12.47 &  11.1$\pm1.7$& 5   \\
  2MASS J08251968+2115521 & L7.5 &  15.12 &  13.79 &  13.05 &  15.68 &  13.91 &  10.7$\pm0.1$& D02, 6 \\
  2MASS J08354256-0819237 & L5 &  13.15 &  11.95 &  11.16 &  13.74 &  12.02 &  8.3$\pm1.2$& 5    \\
  2MASS J08472872-1532372 & L2 &  13.52 &  12.63 &  12.05 &  14.07 &  12.76 &  17.5$\pm2.6$& 5   \\
  2MASS J09083803+5032088 & L7 &  14.56 &  13.47 &  12.92 &  15.14 &  13.62 &  15.9$\pm2.4$& 5   \\
  2MASS J09111297+7401081 & L0 &  12.92 &  12.20 &  11.75 &  13.40 &  12.32 &  17.3$\pm2.6$& 4   \\
  2MASS J09211410-2104446 & L2 &  12.78 &  12.15 &  11.69 &  13.35 &  12.34 &  12.4$\pm1.8$& 7   \\
  2MASS J10452400-0149576 & L1 &  13.13 &  12.37 &  11.81 &  13.69 &  12.45 &  16.8$\pm2.5$& 8  \\
  2MASS J10484281+0111580 & L1 &  12.92 &  12.14 &  11.62 &  13.40 &  12.23 &  15.3$\pm2.3$& 4, 9, 10  \\
  2MASS J10511900+5613086 & L2 &  13.24 &  12.42 &  11.90 &  13.80 &  12.56 &  15.4$\pm2.3$& 4   \\
  2MASS J11040127+1959217 & L4 &  14.46 &  13.48 &  12.98 &  15.09 &  13.64 &  18.8$\pm2.8$&  5 \\
  2MASS J11083081+6830169 & L0.5 &  13.14 &  12.23 &  11.60 &  13.67 &  12.26 &  18.0$\pm2.7$& 8   \\
  2MASS J12130336-0432437 & L5 &  14.67 &  13.68 &  13.00 &  15.29 &  13.77 &  16.7$\pm2.5$& 5  \\
  2MASS J12212770+0257198 & L0 &  13.17 &  12.41 &  11.95 &  13.70 &  12.47 &  19.4$\pm2.9$& 4   \\
  2MASS J14283132+5923354 & L5 &  14.78 &  13.88 &  13.27 &  15.45 &  13.95 &  17.6$\pm2.6$& 4   \\
  2MASS J14482563+1031590 & L5 &  14.56 &  13.43 &  12.68 &  15.21 &  13.57 &  15.9$\pm2.4$ & 4   \\
  2MASS J15074769-1627386 & L5 &  12.82 &  11.90 &  11.30 &  13.44 &  12.05 &  7.34$\pm0.03$& D02, 1 \\
2MASS J15394189-05200428 & L3.5 & 13.92 & 13.06 & 12.58 & 14.61 & 13.16 & 16.2$\pm2.5$ & 10 \\
  2MASS J15525906+2948485 & L1 &  13.48 & 12.61 & 12.03 & 13.60 & 12.02 & 19.8$\pm3.0$ & 2   \\
  2MASS J16580380+7027015 & L1 &  13.31 &  12.54 &  11.92 &  13.83 &  12.57 &  18.6$\pm0.3$& D02, 8  \\
  2MASS J17054834-0516462 & L0.5 & 13.31 & 12.54 & 12.03 & 14.00 & 12.58 & 19.5$\pm2.9$ & 4, 10 \\
  2MASS J17312974+2721233 & L0 &  12.09 & 11.39 & 10.91 & 12.73 & 11.43 & 11.8$\pm1.8$ & 4 \\
  2MASS J17534518-6559559 & L4 & 14.10 &  13.11 &  12.42 &  14.76 &  13.21 &  15.9$\pm2.4$& 4   \\
  2MASS J18071593+5015316 & L1.5 & 12.96 & 12.15 &  11.61 &  13.49 &  12.25 &  14.6$\pm2.2$&  5  \\
  2MASS J19360262-5502367 & L4 & 14.49 &  13.63 &  13.05 &  15.11 &  13.71 &  15.4$\pm2.3$& 7   \\
  2MASS J20575409-0252302 & L1.5 & 13.12 & 12.27 & 11.75 & 13.81 & 12.31 & 15.7$\pm2.2$ & 11 \\
  2MASS J21041491-1037369&  L2.5 & 13.84 & 12.96 & 12.36 & 14.59 & 13.01 & 18.7$\pm2.7$ & 5 \\
  2MASS J22244381-0158521 & L4.5 &  14.05 &  12.80 &  12.01 &  14.71 &  12.95 &  11.4$\pm0.1$& 6 \\
  2MASS J23254530+4251488 & L7.0 &  15.51 &  14.46 &  13.81 &  16.17 &  14.54 &  16.3$\pm2.4$& 12   \\
  2MASS J23515044-2537367 & L0.5 &  12.46 &  11.73 &  11.29 &  12.90 &  11.84 &  13.2$\pm2.0$&12  \\

\enddata
\tablecomments{ D02: Trigonometric parallax from Dahn et al (2002) \\
V04: Trigonometric parallax from Vrba et al (2004) \\
All other distances are based on spectroscopic parallaxes, using the Cruz et al (2003) 
(M$_J$, sp. type) calibration; 
the cited distance uncertainties correspond to an uncertainty of $\sim0.5$ in spectral class. \\
Discovery papers:
1: Reid et al, 2001; 2: Wilson et al, 2003; 3. Geballe et al, 2002; 4. Reid et al, in prep.; 
5. Cruz et al, 2003; 6: Kirkpatrick et al, 2000; 7: Deacon, Hambly \& Cooke, 2005; 8: Gizis, 2002; 
9. Hawley et al, 2002; 10. Kendall et al, 2004; 11: M\'enard et al, 2002; 12: Cruz et al, 2006\\
Individual objects: \\
2M1048+01: Kendall et al cite a spectral type of M7, based on IR spectra; our optical spectrum
indicates a type of L1, in agreement with Hawley et al. \\
2M1707-05 has H$\alpha$ emission and may have weak Li absorption (see Schmidt \& Cruz, 2006, for further details. \\ 
}
\end{deluxetable}

\begin{deluxetable}{rrrrrrrrrrc}
\tabletypesize{\scriptsize}
\tablecolumns{11}
\tablenum{2}
\tablewidth{0pt}
\tablecaption{Observational data for candidate binaries}
\tablehead{\colhead{2MASS name} & \colhead{Sp. type} &
\colhead {J} &  \colhead {H} &\colhead {K$_S$} & \colhead{m$_{110}$} & \colhead{m$_{170}$}& \colhead{d (pc.)} & \colhead {$\delta$ arcsec.} &
\colhead{$\theta$} & \colhead {Notes }}
\startdata
2MASS J00043484-4044058 & L4.5 & 13.11 & 12.06 & 11.40 & & & 15$\pm3$ & 0.09& &LHS 102, 1 \\
A & L4.5 & 13.82 & 12.76 & 12.10 & 14.5 & 12.8\\
B & L4.5 &  13.90 & 12.85 & 12.20 & 14.6 & 12.9\\
2MASS J00250365+4759191 & L5 &14.86 & 13.65 & 12.91 &  &  & 31$\pm6$ & 0.33 & -126.9 & 2, Li\\
A & L4 &15.51 & 14.30 & 13.55 & 16.23 & 14.25 & \\
B & L4 &15.72 & 14.47 & 13.75 & 16.40 & 14.36 & \\
2MASS J01473282-4954478 & M8 & 13.06 & 12.37 & 11.92 &  & & 33$\pm6$ & 0.19 & 67.6 &  3\\
A & M8: & 13.35 & 12.67 & 12.22 & 13.54 & 13.61\\
B & L2: & 14.70 & 14.00 & 13.60 & 14.90 & 14.11 \\
2MASS J04291842-3123568 & M7.5 & 10.89 & 10.21 & 9.80 &  &  & 11.5$\pm2.3$ & 0.55 & -73.0& 4, 5  \\
A & M7.5 & 11.18 & 10.55 & 10.14 & 11.55 & 10.53 &  &&& \\
B & L1: & 12.38 & 11.65 & 11.12 & 12.70 & 11.56 & &&&  \\
2MASS J07003664+3157266 & L3.5 & 12.92 & 10.96 & 11.31 & & & 12.2$\pm4.0$ & 0.17 & 102.0& 6\\
A & L3.5 & 13.23 & 12.27 & 11.62 & 13.48 & 11.58 & & \\
B & L6: & 14.40 & 13.45 & 12.85 & 14.70 & 12.78 & & \\
2MASS J09153413+0422045 & L7 & 14.55 & 13.53 & 13.01 &  & & 14.8$\pm3.0$ & 0.73 & -155.0& 3 \\
A & L7 & 15.30 & 14.28 & 13.75 & 15.99 & 14.28 & \\
B & L7 & 15.40 & 14.40 & 13.85 & 16.11 & 14.37 & \\
2MASS J17072343-0558249 & L1 & 12.05 & 11.26 & 10.71 & & & 15.0$\pm3.0$ & 1.00 & 35.1& 3 \\
A & M9 & 12.25 & 11.46 & 10.90 & 12.78 & 11.53 & \\
B & L3 & 14.00 & 12.90 & 12.4: & 14.56 & 12.84 & \\
2MASS J21522609+0937575& L6: & 15.19 & 14.08 & 13.34 &  & & 24.2$\pm5.0$ & 0.25 & 105.5& 3 \\
 A & L6: & 15.95 & 14.80 & 14.10 & 16.75 & 14.75 & \\
 B & L6: & 16.00 & 14.85 & 14.15 & 16.80 & 14.80 & \\
2MASS J22521073-1730134 & L7.5: & 14.31 & 13.36 & 12.90 &  & & 14.3$\pm3.0$ &0.14 & -18.5 & 3, 7 \\
A & L6: & 14.67 & 13.62 &  &15.43 & 13.69 & &&& \\ 
B & T2: & 15.65 & 15.20 & & 16.55 & 15.25 & &&& \\
\enddata
\tablecomments{ \\ 
Most J \& H magnitudes for the individual components based on deconvolving the 2MASS data using the HST flux ratios, with the K$_S$-band data based on the average colors for the spectral type. The exception is LHS 102, where the data are from Golimowski et al (2004b). \\
Distances estimates are computed from the (M$_J$, spectral type) relation (Cruz et al, 2003), using
data for the primary, except for LHS 102 (parallax estimate by Golimowski et al, 2004b) and 
2M0700+31 (trigonometric parallax by Tinney, Burgasser \& Kirkpatrick, 2003). We adopt 
uncertainties of $\pm20\%$ for the spectroscopic parallaxes \\. 
References: \\
1. Golimowski et al, 2004b; 2. Cruz et al, 2006; 3. Reid et al, in prep.; 4. Cruz et al, 2003; 5. Siegler et al, 2005; 
6. Thorstensen et al, 2003; 7. Reid et al, 2006.}
\end{deluxetable}

\begin{deluxetable}{rrrccrrrrc}
\tabletypesize{\scriptsize}
\tablecolumns{8}
\tablenum{3}
\tablewidth{0pt}
\tablecaption{Inferred properties of binary components}
\tablehead{\colhead{name} & \colhead{M$_J$} &
\colhead {M$_{bol}$} &  \colhead {$M$(1 Gyr)} &\colhead {$M$(3 Gyr)} & \colhead{$q$(1 Gyr)}& \colhead{$q$(3 Gyr)} \\
\colhead{ } & \colhead {} & \colhead {} & \colhead {$M_\odot$} & \colhead {$M_\odot$} & }
\startdata
LHS 102B & 12.95 & 14.7 & 0.070 & 0.078 & 0.99 & 1.00 & \\
LHS 102C & 13.00 & 14.8 & 0.069 & 0.078 & \\
2M0025+4759A & 13.05 & 14.9 & 0.048$^*$ &  & 0.99 & & \\
2M0025+4759B & 13.25 & 15.1 & 0.047$^*$ &  & \\
2M0147-4954A & 10.75 & 12.8 & 0.085 & 0.084 & 0.88 & 0.95 & \\
2M0147-4954B& 12.10 & 14.0 & 0.075 & 0.080 & \\
2M0429-3123A & 10.90 & 12.9 & 0.086 & 0.084 & 0.87 & 0.95 & \\
2M0429-3123B & 12.10 & 14.1 & 0.075 & 0.080 & \\
2M0700+3157A & 12.80 & 14.5 & 0.071 & 0.079 & 0.85 & 0.95 \\
2M0700+3157B & 14.00 & 15.5 & 0.060 & 0.075 & \\
2M0915+0422A & 14.45 & 15.9 & 0.052 & 0.072 & 1.0 & 1.0 \\
2M0915+0422B & 14.45 & 15.9 & 0.052 & 0.072 & \\
2M1707-0558A & 11.35 & 13.3 & 0.081 & 0.082 & 0.89 & 0.94 & \\
2M1707-0558B & 13.10 & 14.9 & 0.072 & 0.077 & \\
2M2152+0937A & 14.00 & 15.5 & 0.060 & 0.075 & 1.0 & 1.0 & \\
2M2152+0937B & 14.05 & 15.55 & 0.060 & 0.075 & \\
2M2252-1730A & 13.90 & 15.4 & 0.061 & 0.075 & 0.66 & 0.87 & \\ 
2M2252-1730B & 14.90 & 16.6 & 0.040 & 0.065 & 
\enddata
\tablecomments{ \\ 
Bolometric corrections are based on data from Golimowski et al (2004a), and
masses are from the Burrows et al (1997) set of theoretical models.\\
$^*$ The presence of strong lithium absorption in 2M0025+4759 implies that the age is less than 1 Gyr, and the masses and mass ratio listed are for an age of 0.5 Gyrs.
}
\end{deluxetable}

\begin{deluxetable}{rrrrrrccrrc}
\tabletypesize{\scriptsize}
\tablecolumns{11}
\tablenum{4}
\tablewidth{0pt}
\tablecaption{Additional L dwarf binaries within 20 pc}
\tablehead{\colhead{2MASS name} & \colhead{Sp. type} & \colhead{d (pc.)} & \colhead {$\delta$ arcsec.} & \colhead{M$_J$} &
\colhead {M$_{bol}$} &  \colhead {$M$(1 Gyr)} &\colhead {$M$(3 Gyr)} & \colhead{$q$(1 Gyr)}& \colhead{$q$(3 Gyr)} & \colhead {Notes } \\
\colhead{ } & \colhead {} & \colhead {} & \colhead {} & \colhead {} & \colhead {} & \colhead {$M_\odot$} & \colhead {$M_\odot$} & }
\startdata
2MASS J02052940-1159296 & L7 & 19.8 & 0.51 & 13.10 & 14.6 & & & 1.0 & 1.0 & 1\\
A & L7 & & & 13.85& 15.35& 0.062 & 0.075& \\
B & L7 & & & 13.85& 15.35& 0.062 & 0.075 & \\
2MASS J04234857-0414035 & L7.5/T0 & 15.2 & 0.16 & 13.54 & 15.0 & & &0.85 & 0.91 & 2 \\
A & L6 & & & 14.05 & 15.55 & 0.060 & 0.078 &\\
B & T2 & & & 14.65 & 15.95 & 0.051 & 0.071 & \\
2MASS J07464256+2000321 & L0.5 & 12.2 & 0.22 & 11.31 & 13.25 &  & &0.97 & 0.98 & 3 \\ 
A & L0.5 & & & 11.9 & 13.85 & 0.076 & 0.081  &\\
B & L2 & & & 12.3 & 14.15 & 0.074 &0.079  & \\
2MASS J1305401-254106 & L2 & 18.66 & $<$0.3 & 12.06 & 13.65 &  & & 0.90$^*$ & & 4 \\  
A & $\sim$L1.5 & & & 12.75 & 14.35 & 0.050$^*$ & \\
B &  $\sim$L4 & & & 13.25 & 14.75 & 0.045$^*$ & \\
\enddata
\tablecomments{ Notes:\\
1. Denis-P J0205.4-1159, see Koerner et al, 2001 \\
2. SDSS J042348.57-041403.5, see Burgasser et al, 2005 \\
3. see Reid et al, 2001 \\
4. Kelu 1, see Liu \& Leggett, 2005 \\
Bolometric corrections based on the Golimowski et al (2004a) BC$_K$/spectral-type relation; \\
Masses from mass-luminosity relations in Burrows et al (1997), except Kelu 1, where the mass ratios are from Liu \& Leggett (2005) for $\tau=300$ Myrs.\\ 
}
\end{deluxetable}

\centerline {Figure captions}

\figcaption{NIC1 F110W images of candidate binaries from in Table 1; the x and y axes are in pixels (0.043 arcseconds per pixel) and the arrows indicate the orientation on the sky. Images of 2M2252-1730 are include in RLBC06, while Golimowski et al (2004b) present higher-resolution ACS images of LHS 102 BC. In addition to the seven binary candidates, we include an image of 2M1705-0516, showing the possible companion. Further details on the binary systems are given in Table 2.}

\figcaption{ J-band bolometric corrections as a function of spectral type. The bolometric magnitudes are taken from the multiwavelength analysis y Golimowski et al (2004a), while the J-band data are from 2MASS}

\figcaption{The NIC1 point-spread functions in the F110W and F170M filters: the upper panels plot the psf in linear units (counts sec$^{-1}$ pix$^{-1}$) for 2M0523-1403 (J=13.12, H=12.22, L2.5); the lower panels plot the extended psf in logarithmic units. The dashed lines in the lower panels mark the effective detection limit for isolated point sources (see \S4.1 for further discussion).}

\figcaption{ The dotted lines plot (M$_{bol}$, mass) isochrones from the Burrows et al (1997) models for ages of 0.3, 0.5, 1, 3 and 5 Gyrs (from left to right), while the dashed lines plot data for the Chabrier et al (2000) "dusty" models for ages 0.5, 1 and 5 Gyrs. The horizontal lines mark the location of the binary components listed in Tables 2 and 3; the vertical bar on the M$_{bol}=14.9$ line marks the 0.065$M_\odot$ limit for 2M0025+4759A.}

\figcaption{ The age distribution of early-type and late-type L dwarfs predicted by Allen et al (2005); the data are plotted as a cumulative distribution, with the ordinate marking the probability that a field dwarf has an age younger than the value of the abscissa. The solid line plots the predictions for L0 dwarfs, the dashed line for L5 dwarfs and the dot-dashed line for L6-L8 dwarfs.}

\figcaption{ The F110W psf for 2M0523-1403 plotted in a magnitude scale, relative to the peak brightness. The solid points mark the location of nine probable binaries listed in Table 2, and the 5-point stars plot data for the other nearby binaries listed in Table 4. The dotted lines mark the range of effective detection limits set by the sky background level of the present NICMOS observations. The sequence of short horizontal lines indicates the magnitude difference for a 0.07 $M_\odot$ primary and a secondary with $0.8 \ge q \ge 0.2$ and ages 0.5, 1.0 and 5 Gyrs.}

\figcaption{The posterior probability distribution derived for the four-parameter model outlined in \S4.3, based on Bayesian analysis of the current ultracool sample. Panel a) shows the overall binary frequency; panel b) shows the mean value of the semi-major axis distribution (in AU); panel c) plots the probability distribution of the width of the Gaussian semi-major axis distribution (in logarithmic units); and panel d) plots power-law index, $\gamma$, derived for the mass-ratio distribution. The best-fit values derived for each parameter are listed in the text.}

\figcaption{A comparison between the results derived in this paper for the L dwarf mass-ratio and semi-major axis distribution, and results for M dwarfs and G dwarfs. The solid histogram plots the observed L dwarf distributions, and the dotted histogram shows the resulting best-fit model. The solid squares in the upper and lower panels mark results derived by Fischer \& Marcy (1992) for M dwarfs; note that they derive a high spectroscopic binary fraction ($\sim$0.6 on this scale), but with a substantial ($\pm60\%$) uncertainty. The open triangles in the upper panel and the dashed line in the lower panel (a log-normal distribution with $\langle a \rangle = 1.60$, or 40 AU) plot results deduced by Duquennoy \& Mayor (1991) for all G dwarf binaries, while the crosses in the upper panel plot the mass ratio distribution derive by Mazeh et al (1992) for spectroscopic G-dwarf binaries; clearly, G dwarfs have a much higher proportion of low mass-ratio systems and a broader semi-major axis distribution.}

\newpage
\plotone{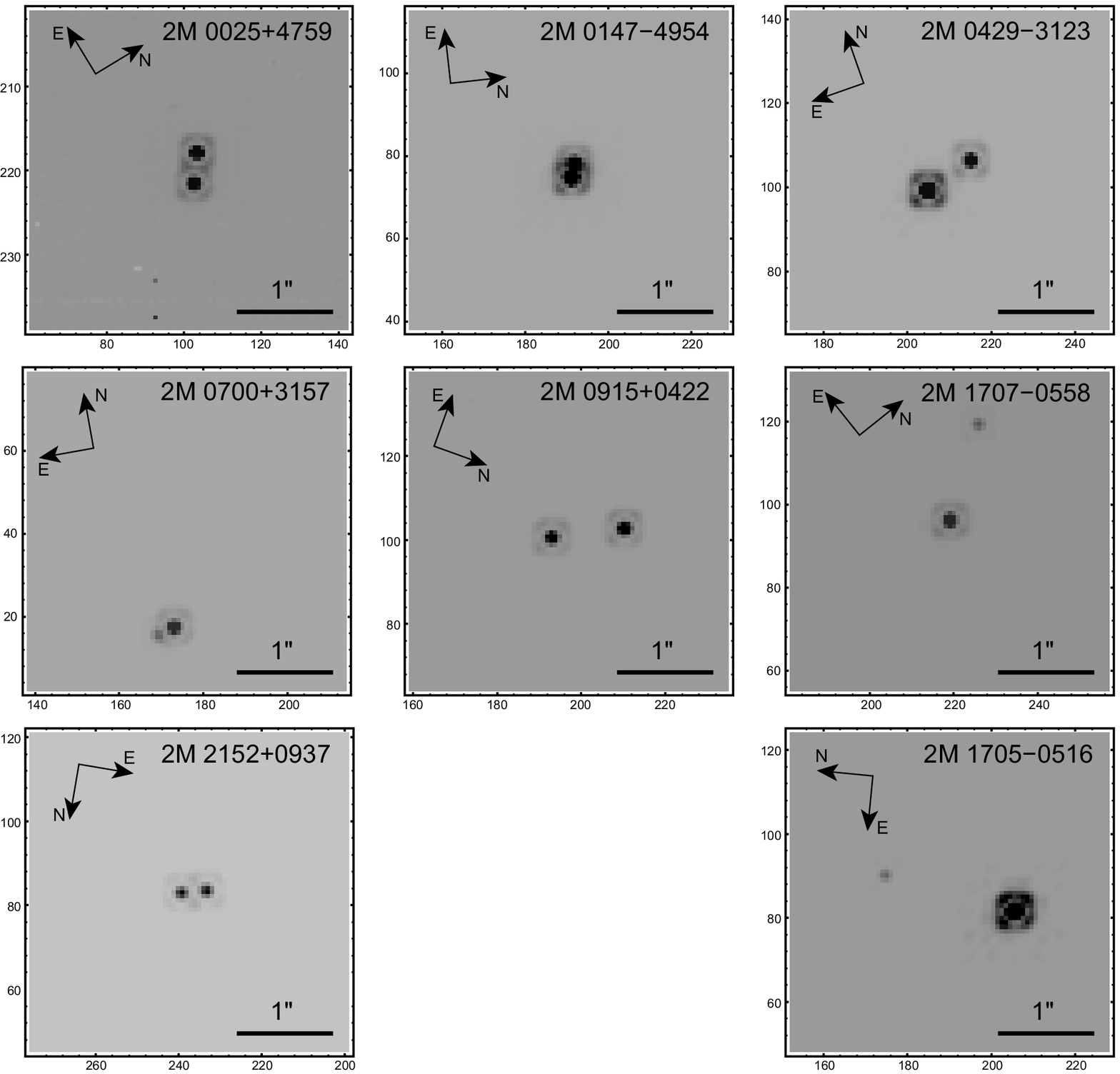}
\newpage
\plotone{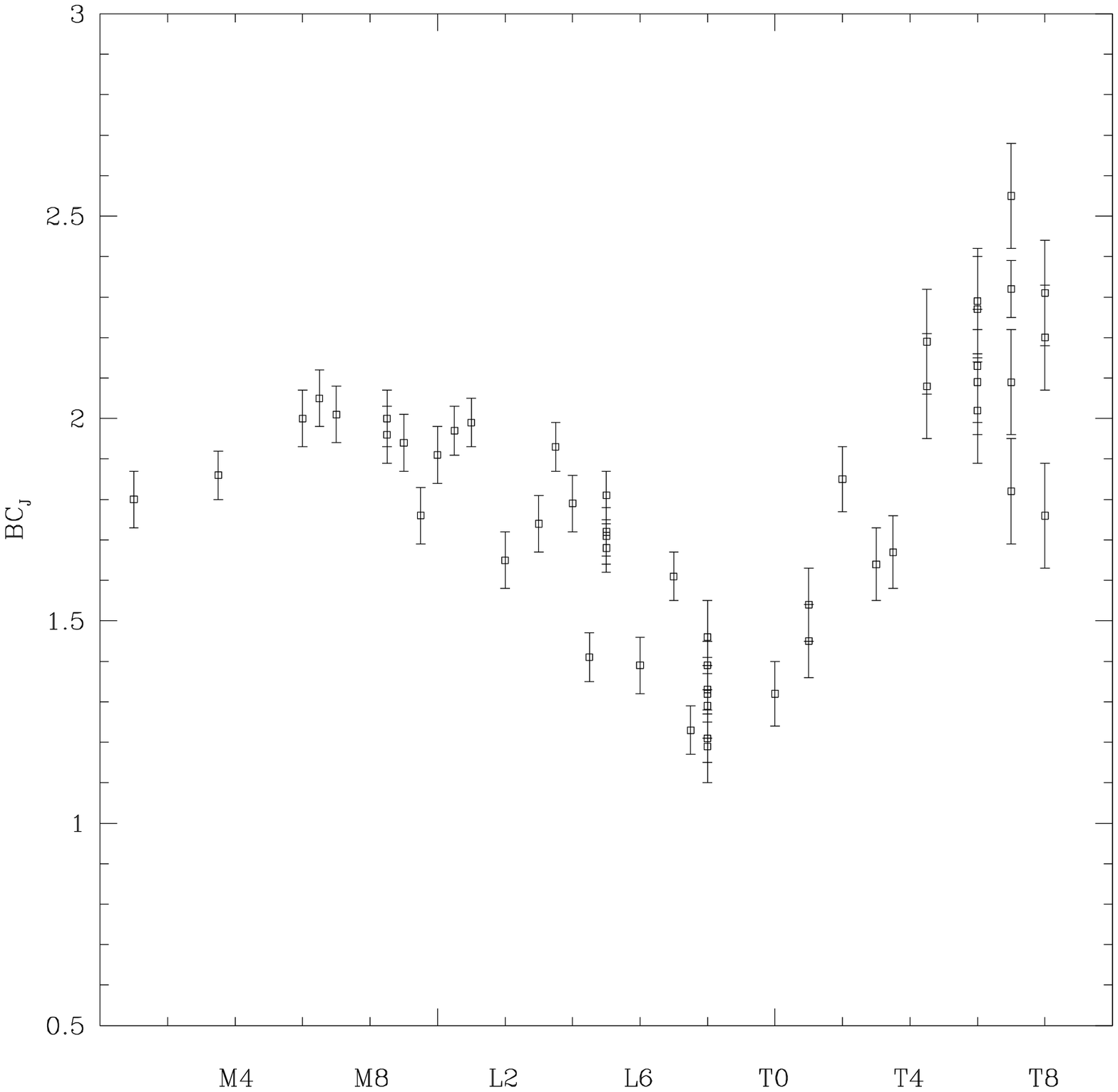}
\newpage
\plotone{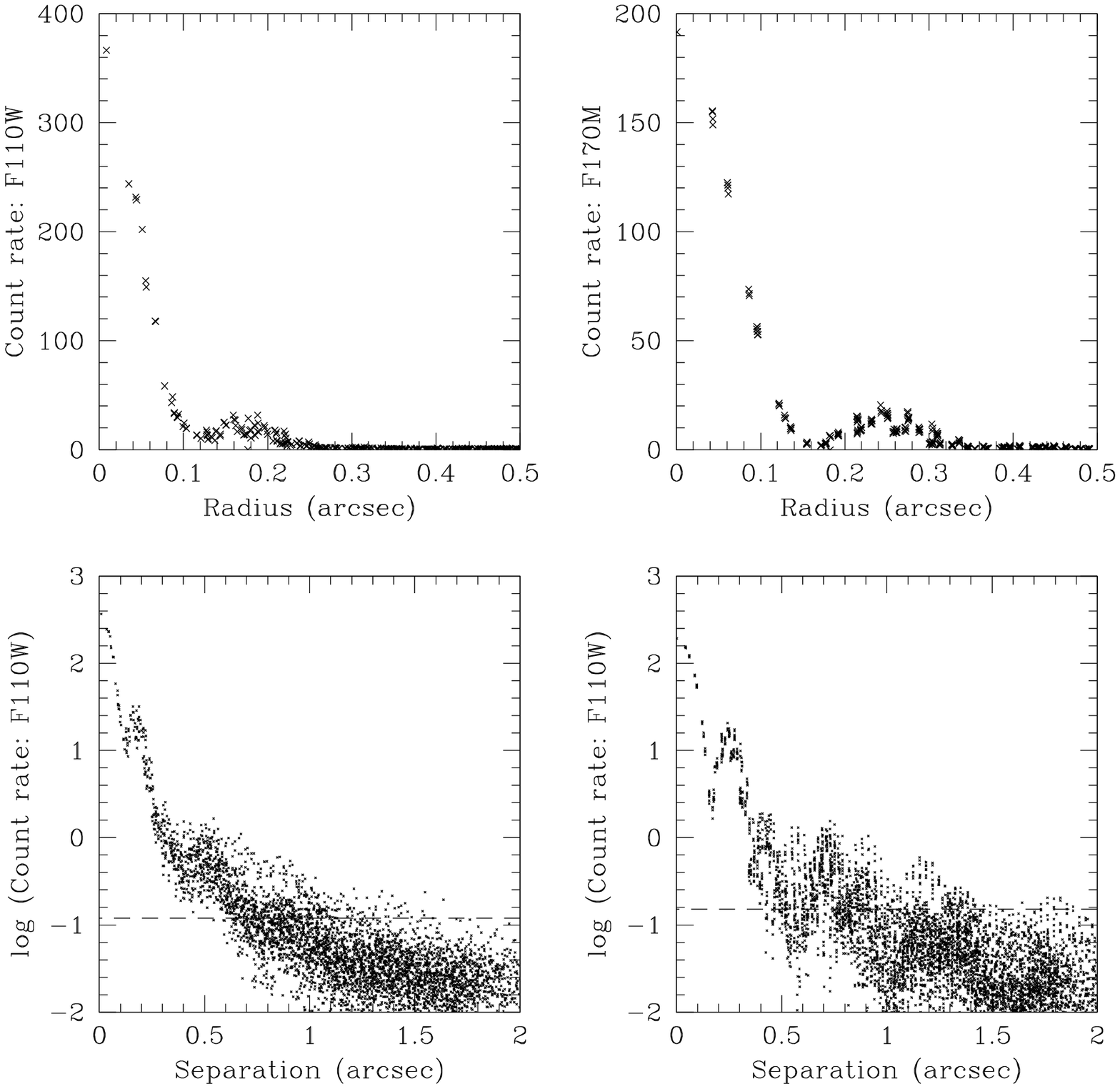}
\newpage
\plotone{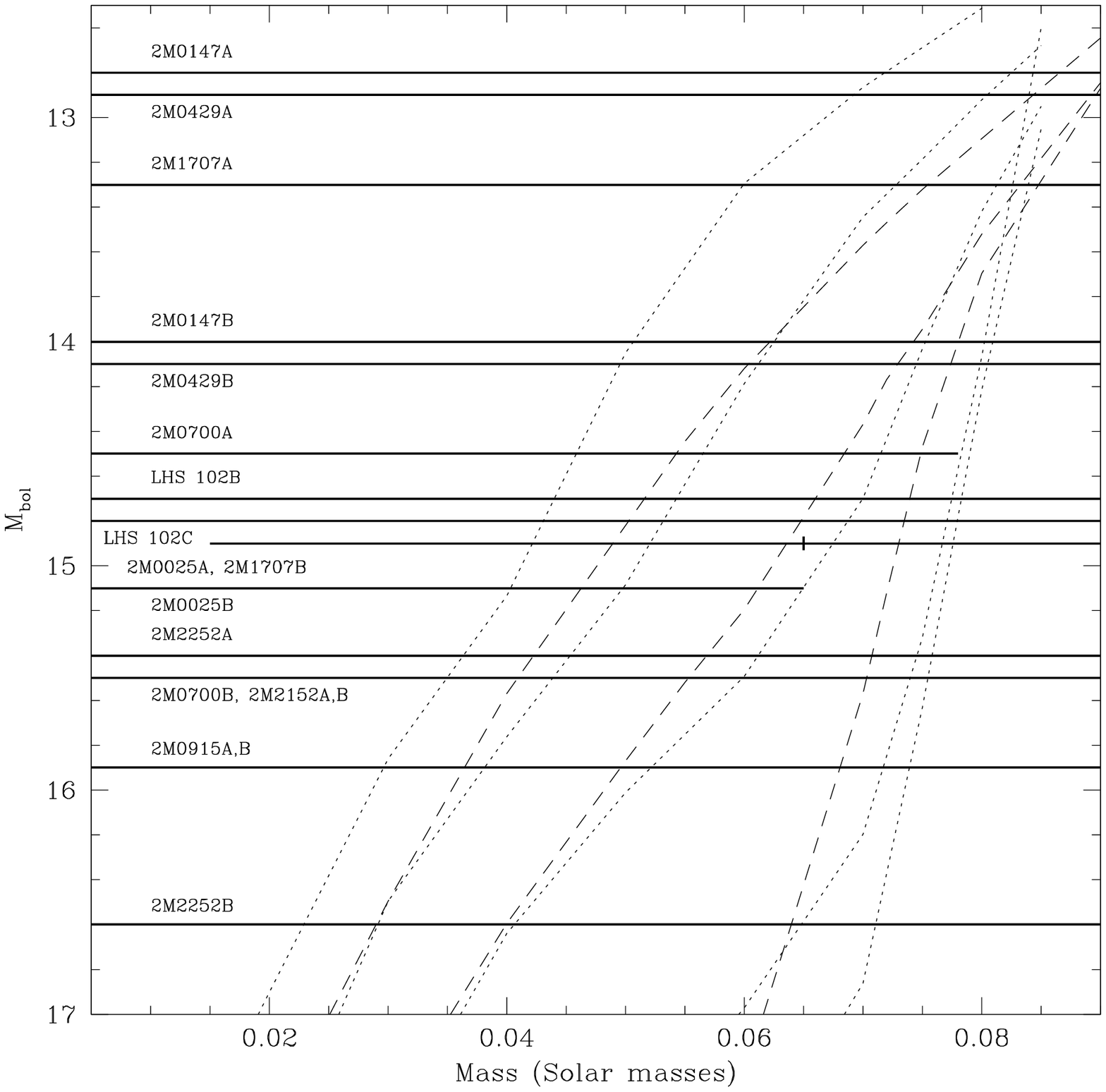}
\newpage
\centerline{
\includegraphics[angle=-90,scale=.9]{reid.f5.eps}}
\newpage
\plotone{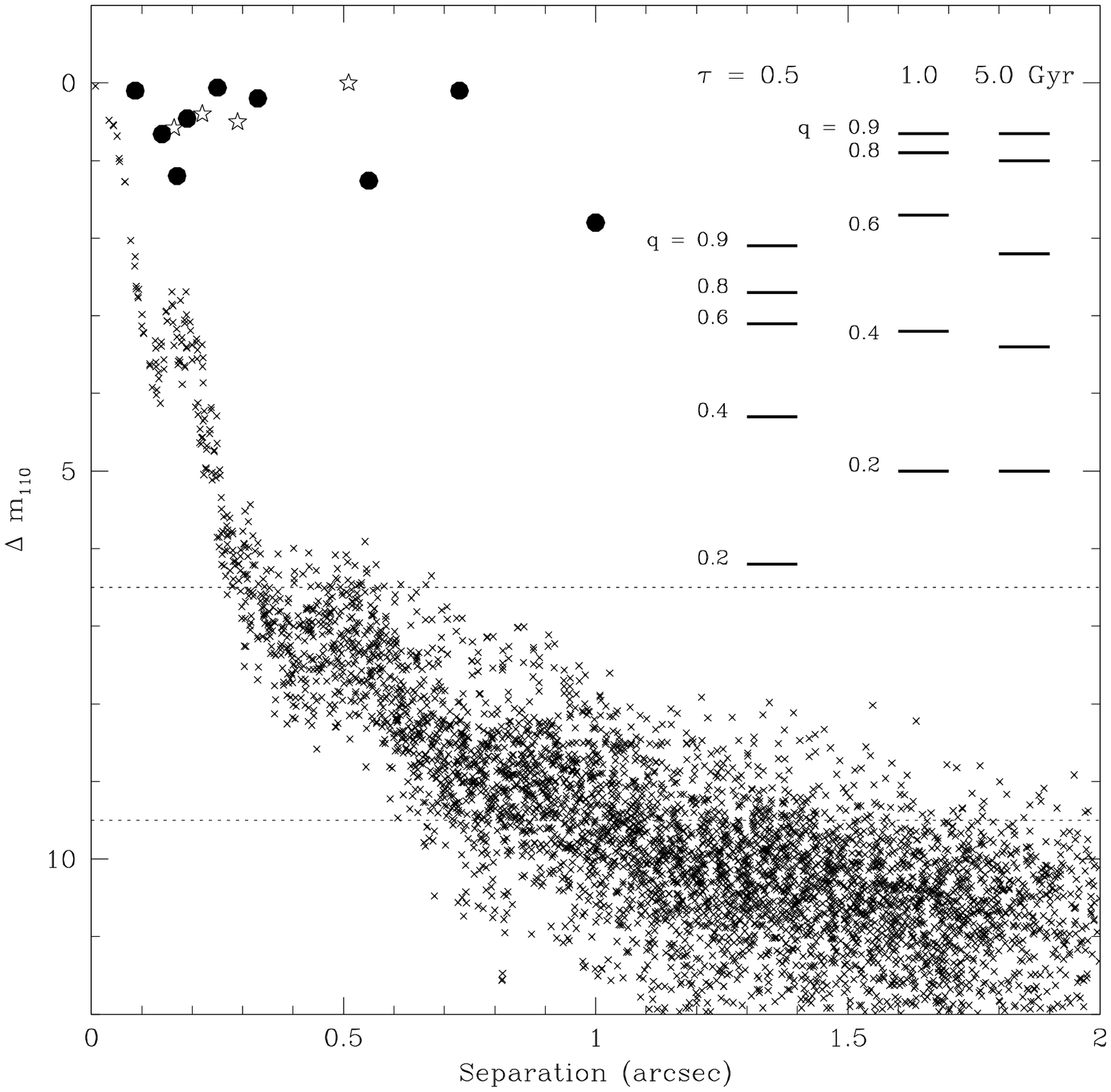}
\newpage
\centerline{
\includegraphics[angle=-90,scale=1.4]{reid.f7.eps}}
\newpage
\plotone{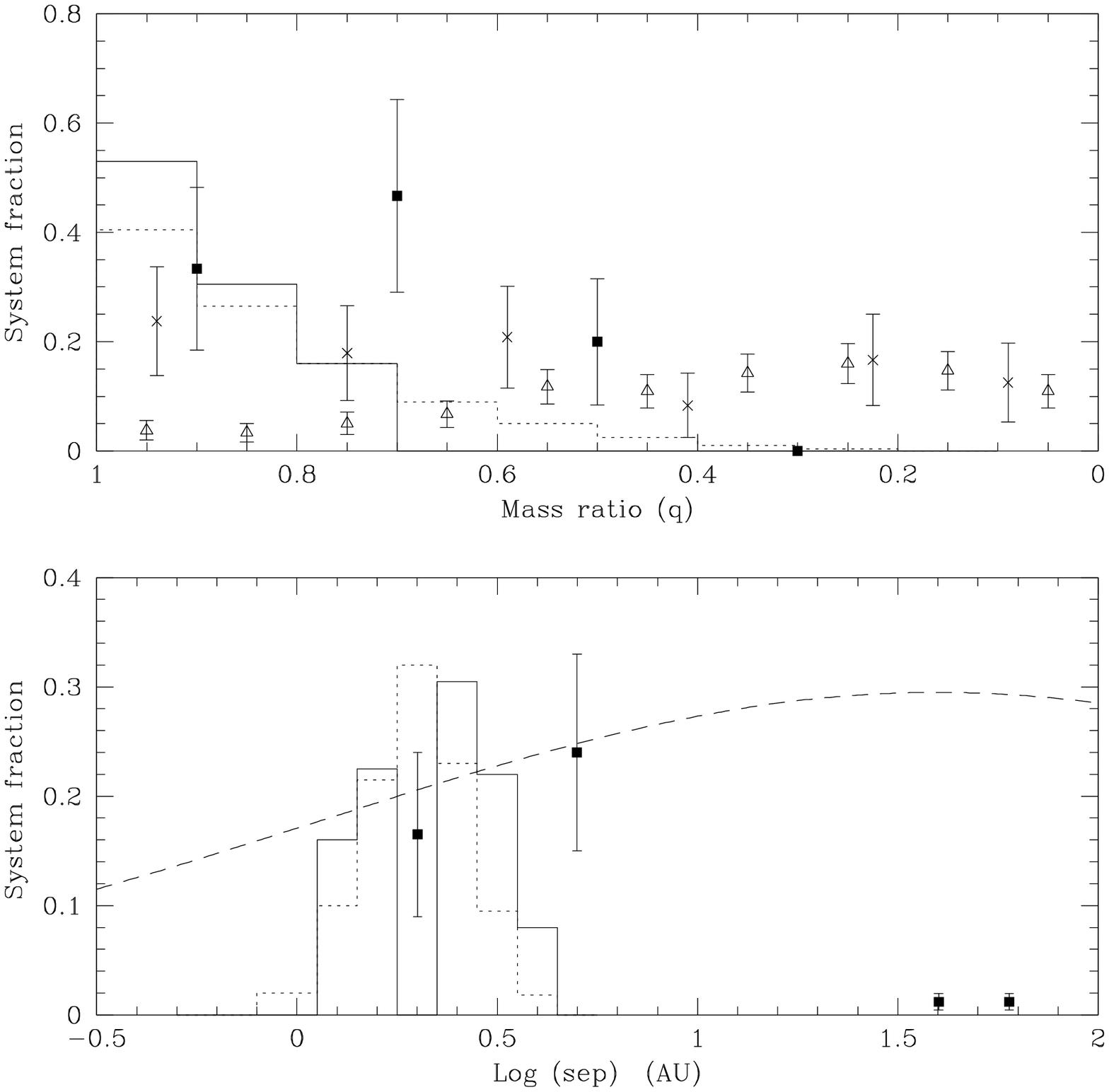}

\end{document}